********** tex file begins *********
\documentstyle[preprint,aps]{revtex}
\begin{document}
\draft
\tightenlines
\title{
\bf{ 
Infinite number of exponents for a spin glass transition
} }
\author{Sutapa Mukherji\cite{eml1} and Somendra M.
Bhattacharjee\cite{eml1}} 
\address{Institut f\"ur Festk\"orperforschung, Forschungzentrum,
D 52425 J\"ulich, Germany\\
Institute of Physics,
Bhubaneswar 751 005, India}
\maketitle
\begin{abstract}
We consider the behavior of the overlap of $m (\geq 2)$ paths at the
spin glass transition for a directed polymer in a random medium.
We show that an infinite number of exponents is required to
describe these overlaps.  This is done in an $\epsilon = d-2$
expansion without using the replica trick. 
\end{abstract}
\pacs{75.10.Nr, 75.40.Cx, 36.20.Ey, 05.70.Fh }

When disorder induces a new thermodynamic phase not found in the pure
system, the description of the transition itself poses new challenges.  This
is exemplified by the efforts made to understand the spin glass transition
\cite{mpv,fischer}.  Other examples are magnets in random fields\cite{randh}, 
polymers in disordered media \cite{bkc,kpz,tang,halp} or with random
interactions \cite{sml}, etc. Several concepts emerged from the solution of
the infinite range ``mean field" spin glass problem using the replica trick,
but their validity for finite dimensional systems, remains to be
established\cite{fishu,moore}.

The most important concept in the replica approach is that of the overlap
which plays the role of the order parameter for the transition\cite{mpv}.
The overlap purports to characterize the rugged free energy landscape in the
spin glass phase through the statistics of the pairwise common
configurations in the various minima.  In principle, a many valley free
energy surface would require higher order overlaps for a more detailed
description\cite{derphy,code}.  These are the overlap of say, three or more
($m$) minima, to be called ``{\it m-overlap}".  All of these overlaps vanish
at the transition point as $q_m \sim \mid T_c - T\mid ^{\beta_m}$ for $T
\rightarrow T_c-$,because the free energy surface goes over to a smooth one
in the high temperature phase.  Now, how many exponents are needed to
describe these overlaps?  The answer is one for {\it infinite range or
infinite dimensional} models \cite{mpv,derphy}, justifying the use of the
pair overlap, $q_2$, as the sole order parameter in the replica approach
(with $\beta_2$ as the order parameter exponent)\cite{comm0,comm1}.  What
about finite dimensional systems? The question assumes importance because
the number of independent exponents tells us the number of quantities one
requires to characterize the transition.  Alas, so little is known for the
spin glass problem \cite{fischer}.

In this paper, we develop a method to calculate these $m$-overlaps for
arbitrary $m$ in an $\epsilon = d-2$ expansion for the spin glass transition
of a directed polymer in a random medium (DPRM) {\it without using the
replica trick}. A $d+1$ dimensional DP is a string stretched in a preferred
direction and with free fluctuations in the transverse $d$ dimensions.  In a
random medium, the gain in the potential energy from randomness can win over
the ``random walk'' entropy, producing a disorder dominated `super
diffusive'' phase \cite{kpz,tang,halp,mez,parisi}.  For $d>2$, there is a
transition from a low temperature, strong disorder, spin glass type phase to
a pure type phase \cite{tang,der1,doty}.  The transition is described by an
unstable fixed point $(\sim O(\epsilon))$ in a renormalization group (RG)
approach via the mapping \cite{henl} to a nonlinear noisy stochastic
equation for the free energy (Kardar-Parisi-Zhang equation)\cite{kpz,tang}.
The simplicity of the model, nontrivial solutions in many
cases\cite{bethe,derphy,code}, and the possibility of studying various
fundamental questions\cite{medkar,scar} related to disorder systems in
general, make the strong disorder phase and the transition a topic of
paramount importance \cite{halp}.  In fact, various techniques have been
used for this purpose, as, e.g., Bethe ansatz for $d=1$ \cite{bethe},
dynamic renormalization group \cite{fos,kpz,tang,smr}, scaling
theory\cite{doty}, numerical simulations \cite{kost1}, mode
coupling\cite{moorel}, $1/d$ expansion \cite{gold}, on a Cayley tree
\cite{derphy}, on hierarchical lattices \cite{code} etc.  Replica symmetry
breaking \cite{comm} has been tried by some \cite{mez,parisi,blum} and 
vehemently opposed by a few others\cite{hf,hwfis}.

For DP, the overlap ($m$-overlap) describes the fraction of paths common to
two ($m$) minimum (free) energy paths, and is therefore equal to the
fraction of the paths two ($m$) polymers go together when placed in the same
random medium \cite{mez,parisi,derphy,code,smr,palad}. The two ($m$)
polymers act as the replicas of the one chain system.  In the spin glass
phase, if there is only one minimum free energy path, then both the chains
would follow the same path, an attraction induced by the disorder.  In case
there are many valleys, then the chains can get separated by hopping to a
neighboring valley.  Thus overlaps contain information about the valley
structure.

The $m$-overlaps, $q_m$, have been calculated for a DPRM on a Cayley tree
and on hierarchical lattices \cite{code,derphy}.  The Cayley tree problem
can be thought of as an infinite dimensional case while the hierarchical
lattices are definitely finite dimensional with tunable dimensionality.  For
the Cayley tree problem, closed form expressions for $q_m$ show that
$\beta_m=1$ for all $m$ \cite{derphy}.  For hierarchical lattices, there is
a critical dimension above which a transition takes place \cite{code}.
Numerically the transition temperature has been obtained by locating the
temperature where $q_2$ and $q_3$ vanish.  Nothing, unfortunately, is known
for $\beta_m$.  A linear dependence of $\beta_m$ on $m$, in a multifractal
analysis, would also mean that only one exponent is needed, as e.g.  for
pure noninteracting Gaussian chains, (see below).  The crucial question is,
therefore, whether such linearity is maintained for finite dimensional
systems.  The answer we find is no.

In this paper, we use the continuum approach. The polymer is described by
the Hamiltonian
\begin{equation}
 H=\int\limits_{0}^{t} d\tau \ \left [ \frac{\gamma}{2}
\ {\bf {\dot x}}^2(\tau)+ \frac{\lambda}{2\gamma}
V({\bf x}(\tau),\tau)\right ]
\label{eq:onepoly}
\end{equation}
where ${\bf x}(t)$ is the $d$ dimensional transverse spatial coordinate of
the DP at the contour length $t$, and ${\bf {\dot x}}(\tau) =d{\bf
x}(\tau)/d\tau$.  The first term on the right hand side represents the
entropic fluctuations of a free Gaussian chain with $\gamma$ as the bare
line tension. $V$ corresponds to a space and time dependent random potential
seen by the chain, and the amplitude ${\lambda}/{2\gamma}$ is chosen for
convenience.  The random potential is taken to be uncorrelated, normally
distributed \cite{kpz} with zero mean and $\overline{V({\bf x},\tau)V({\bf
y},\tau ')}= 2 \Delta\ \delta({\bf x} -{\bf y})\delta(\tau-\tau')$ where the
overbar stands for disorder averaging.

A formal way to define the $m$-overlap is to put $m$ chains in the system
and take
\begin{equation}
q_m=-\frac{1}{t} \int_0^t d\tau \overline{\left\langle
\prod_{i=1}^{m-1}\delta{\bf (x}_{i,i+1}(\tau){\bf )} 
\right\rangle}, \label{eq:qm}
\end{equation}
where ${\bf x}_{i,i+1}(\tau) = {\bf x}_{i}(\tau) - {\bf x}_{i+1}(\tau)$, and
$\langle ...\rangle$ stands for thermal average for a realization.  One way
of computing $q_m$ is to couple the $m$ chains or replicas with a weak
$m$-body interaction.  Then overlap would follow from an appropriate
derivative of the total free energy with respect to the hypothetical
coupling constant (see below).  This procedure was adopted for the overlap
($q_2$) in the numerical work of Mezard in 1+1 dimensions \cite{mez} and by
one of us in a 1-loop RG approach \cite{smr} (see also Ref. \cite{palad}).
We generalize the method of Ref \cite{smr} for $q_m$.  The RG analysis is
geared towards calculating the scaling exponent of the coupling constant.  A
judicious use of finite size scaling, as explained below, then gives us the
exponent $\beta_m$.

With the definition of the $m$-overlap in Eq. \ref{eq:qm}, we consider an
$m$ chain interacting Hamiltonian
\begin{equation}
{\cal H}_m= \sum_{i=1}^m H_i + (\lambda/2\gamma)
v_m \int_0^t d\tau \prod_{i=1}^{m-1} \delta{\bf
(x}_{i,i+1}(\tau){\bf )} .
\label{eq:mpoly}
\end{equation}
where $H_i$ is the Hamiltonian of Eq. \ref{eq:onepoly} for the $i$th
polymer.  Defining the quenched free energy $f_m (v_m,t) =
\overline{\ln Z_m}$ where $Z_m$ is the partition function for ${\cal
H}_m$, the $m$-overlap is obtained as $q_m = - t^{-1} \quad {d
f_m(v_m,t)}/{d v_m}|_{v_m=0}$.

Our interest is in the scaling part of the free energy, $f_m \approx
t^{\chi/z} {\sf f}(v_m t^{-\phi_m /z})$, where $\phi_m$ is the scaling
exponent for $v_m$, and $\chi$ and $z$ are the single chain free energy
fluctuation ($\Delta f \sim t^{\chi/z}$) and dynamic (size $x
\sim t^{1/z}$) exponents \cite{smr,kpz}.  We have verified that, as
expected, there is no change in the single chain exponent.  Taking
derivative then gives $q_m \sim t^{\Sigma_m}$, with $\Sigma_m= (\chi -
\phi_m -z)/z$.  Incidentally, the system is taken to be infinite in
extent in all the $d$ transverse directions and is of length $t$ in
the preferred direction. This form of $q_m$ can, therefore, be treated
as a finite size scaling form \cite{bar}.  Now, the transition takes
place only in the thermodynamic limit of $t\rightarrow \infty$.  In
that limit, for $d>2$, there is a diverging length scale with exponent
$\nu$, $\xi_{\parallel} \sim \mid T-T_c\mid^{-\nu}$, parallel to the
special $t$-like direction \cite{kpz,der1}.  Finite size scaling
suggests a scaling form $q_m = $ $t^{-\beta_m/\nu}
g(t/\xi_{\parallel})$. Therefore, right at the critical point (i.e.,
the unstable fixed point in RG), $q_m \sim t^{-\beta_m/\nu}$.  A
comparison then yields $\beta_m= -\nu \Sigma_m $. Remember, that $\nu$
is strictly $m$ independent.  Our strategy is therefore to calculate
$\phi_m$.

Define $h(\{{\bf x}_j\},t) = (2\gamma/\lambda) \ln Z(\{{\bf x}_j\},t)$,
where $Z(\{{\bf x}_j\},t)$ is the partition function for chains with end
points at $\{{\bf x}_j\}$, all starting at the origin.  This $h$ satisfies
the equation \cite{smr},
\begin{equation}
\frac{\partial\ }{\partial t} h = \sum_{j=1}^{m} \left( \gamma
\nabla_j^2 h + \frac{\lambda}{2} (\nabla_j h)^2 \right) + g_0,
\label{eq:kpzm}
\end{equation}
where $g_0= \sum_{j=1}^{m} V({\bf x}_j,t) + v_m
\prod_{j=1}^{m-1} \delta{\bf ({\bf x}}_{j,j+1}(\tau){\bf )} $.
This equation, though resembling a higher ($md$) dimensional KPZ equation
\cite{kpz}, is actually not so for the peculiar noise term. We prefer this
equation to Eqs. \ref{eq:onepoly} or \ref{eq:mpoly} because an RG can be
implemented with the nonlinear term $\lambda$ as perturbation.  This is
different from a perturbation in the random potential around the Gaussian
chains. Moreover, the equation describes the free energy and averaging $h$
will naturally give the quenched average free energy without any recourse to
the replica trick.

A scaling of ${\bf x} \rightarrow b {\bf x}, t \rightarrow b^z t$, then
shows that, in the absence of nonlinearity $\lambda$, $v_m$ $\rightarrow
b^{z-(m-1) d - \chi} v_m$.  An anomalous part $\eta_m$ would creep in
through renormalization when the nonlinearity $\lambda$ is present. The
crossover exponent is therefore $\phi_m =-[ z-(m-1)d -
\chi + \eta_m]$.  This gives
\begin{equation}
  \beta_m = \nu[d(m-1) + \eta_m]/z. \label{eq:simeta}
\end{equation}
The $m$ dependence, apart from the Gaussian one, therefore comes from
$\eta_m$.  At the Gaussian level the exponents depend linearly on $m$.

The formal solution of Eq. \ref{eq:kpzm}, in the Fourier space $({\bf
k}_j,\omega)$ conjugate to $({\bf x}_j, t)$, can be written as
\begin{eqnarray}
\lefteqn{h(\{{\bf k}_j\},\omega) = G_0(\{{\bf k}_j\},\omega) {\tilde{g_0}}
 -(\lambda/2) G_0(\{{\bf k}_j\},\omega)\times}\nonumber\\ && \int_{\{\cdot
\}} \left (\sum_{j=1}^{m} {\cal P}_j 
\right ) h (\{{\bf p}_j\},\Omega) h(\{{\bf k}_j - {\bf
p}_j\},\omega-\Omega), \label{eq:soln}
\end{eqnarray}
where $G_0(\{{\bf k}_j\},\omega)=[\gamma \sum_j k_j^2 - i
\omega]^{-1}$ represents the bare $m$ particle propagator (Green's
function)(see Fig. 1), and ${\tilde{g_0}}$ is the Fourier transform 
of $g_0$. A shorthand notation is used, viz, ${\cal P}_j ={\bf p}_j
\cdot ({\bf k}_j - {\bf p}_j)$, $\int_{\{\cdot\}} =
(2\pi)^{-md-1} \int \ d\Omega \prod_{i=1}^{m} d{\bf p}_i$, and $\{{\bf
k}_j\}$ to represent all the $m$ ${\bf k}$ vectors.  To tame
possible divergences, we put an upper cutoff $\Lambda$ (=1) for all
$p$ integrals.  This cut off actually comes from a short distance
cutoff in real space.

We now use a dynamic renormalization group approach to determine the
behavior of $v_m$ for large length scales by integrating out fluctuations on
smaller scales. This is based on Eq.  \ref{eq:soln}.  The procedure is well
documented especially for the KPZ equation \cite{kpz} and for the $m=2$ case
of Eq. \ref{eq:kpzm} \cite{smr}.  The basic idea is to (i) integrate out the
fluctuations at the shortest scale by taking a slice $\Lambda e^{-\delta l}
\leq p \leq \Lambda$ from the momentum integral, (ii) absorb it in the coupling
constant, (iii) and rescale all the momenta etc to get back the original
cutoff $\Lambda$.  The resulting changes are then absorbed by renormalizing
the parameters of the problem.  In the limit $\delta l
\rightarrow 0$, these changes are expressed in terms of differential
equations (recursion relations) that tell us the flow of the
parameters as we go to longer length scales.  Special care is needed,
for the problem in hand, to keep track of the momenta indices that get
coupled by the noise.  These interchain connections produce the
necessary anomalous part in the RG equation.  We skip the algebraic
details.

The flow of the disorder is described in terms of the dimensionless coupling
$U=K_d{\lambda^2\Delta}/(2{\gamma^3})$ where $K_d={(2\pi)^{-d}}{S_d}$, $S_d$
being the surface area of the unit $d$ dimensional sphere, and can be found
in Ref. \cite{kpz}.  We have verified that this single chain equation is
recovered from Eq. \ref{eq:soln}, and is independent of $m$.  Let us
recapitulate that at $d=2$, $U$ is marginally relevant.  This leads to a new
critical point for $d>2$.  For $d=2+\epsilon$, the unstable fixed point $U^*
= 2\epsilon$ corresponds to the spin glass transition point, with
$\nu=2/(d-2)$ as the length scale exponent\cite{comm2}.

We concentrate on the renormalization of the coupling $v_m$.  For the long
wavelength, long time limit, the external wavevectors and frequency are
small or zero. In this limit, for arbitrary $m$, the effective coupling
constant to one loop order (see Fig. 1) is given by (suppressing the zero
wave vectors )
\begin{eqnarray}
\lefteqn{v_{mR}= v_m + 8 \left({m\atop 2}\right) \left (- \frac{\lambda}{2}
\right)^{^2}
(2 v_m \Delta)}\nonumber\\ && \times \int_{p,\Omega} p^4 G_0({\bf p},\Omega)
[G_0(-{\bf p},-\Omega)]^2 G_0({\bf p},-{\bf p},0).\label{eq:vmpert}
\end{eqnarray}
The recursion relation for $v_m$ follows from Eq.
\ref{eq:vmpert} as 
\begin{equation}
\frac{d v_{m}}{dl} =  [z- \chi - (m-1)d +
\left({m\atop 2}\right) U ] v_m. \label{eq:vmrec}
\end{equation}

Since our interest is in the crossover exponent for $v_m$ at $v_m=0$, this
first order (in $v_m$), one loop equation is sufficient for us.  Higher
loops will generate higher order terms in $U$ (and hence $\epsilon$).

Using the one-loop fixed point value $U^* = 2\epsilon$ for the transition
point, we find $\eta_m = -m(m-1)\epsilon$, which from Eqs.
\ref{eq:simeta} gives
\begin{equation}
\beta_m = \nu \zeta_c [2(m-1) - (m-1)^2 \epsilon +
O(\epsilon^2)],\label{eq:betfin}
\end{equation}
where $\zeta_c=1/z_c$ is the size exponent at the transition point.  We now
see that the linear relation between $\beta_m$ and $m$ at the Gaussian level
is not respected in the first order. In other words the exponents $\beta_m$
are not linearly dependent on each other, and higher order terms are
expected to make the interdependence more complicated.  Hence the need for
an infinite number of exponents at the transition point.

From the nature of the perturbation series, we see that the change in
exponent to first order in $\epsilon$ is due to the effective two body
interaction induced by the disorder.  The loop in Fig 1 and in Eq.
\ref{eq:vmpert} comes when disorder couples two different chains.
With $\delta$ correlated noise, this means that the two chains are
going through the same point in space.  The purpose of overlap is to
count these.  It is wellknown that two body interaction changes the
reunion behavior of $m$ chains, and each $m$ requires a new exponent
for reunion \cite{hfstat,smj}.  This is the situation here.

Let us now try to connect this result to a replica analysis.  We take $n$
chains in the random medium and average the resulting partition function, or
equivalently, get the effective Hamiltonian for the $n$th moment of the
partition function.  The effect of the disorder is to couple these chains
through a two body interaction \cite{lasskpz}.  The $m$-overlap, then
corresponds to the reunion of $m$ chains out of these $m$, in the limit
$n\rightarrow 0$ (the replica trick).  At the critical point, taking the
chains to be Gaussian (since $\zeta_c = 1/z_c = 1/2$)\cite{doty,tang}, the
reunion of a subset of interacting random walkers can be studied following
Ref. \cite{smj}.  The only difference with Ref \cite{smj} is that the
relevant fixed point is the unstable one, and the $n\rightarrow 0$ limit can
be taken, [see, e. g., Ref. \cite{smj}(b)] to get the anomalous part of Eq.
\ref{eq:betfin}.  This is essentially correct but it still needs to be
established that the chains are actually Gaussian (not just
$\zeta_c=1/2$). These problems are not present in the differential
equation approach used in this paper.

What do all of these mean for the spin glass transition in finite
dimensions?  One expects to to write down (in the $n \rightarrow 0$ limit) a
Landau-Ginzburg type free energy functional with the overlaps as the order
parameters. There can be two possibilities. (A) One is that the simple
minded description through the overlap of two copies is not sufficient, and
one has to worry about the higher overlaps, and in fact an infinite number
of them.  Even if one starts with the Gaussian distribution, renormalization
effects will generate the higher overlaps (arbitrary distributions generate
then in any case).  In this situation, the conventional replica approach may
not be useful. The problem here again may be the interchange of the two
limits, viz,$n\rightarrow 0$ and the thermodynamic limit.  (B) The other
option is that the two copy overlap is good enough in the sense that higher
$m$-overlaps are irrelevant.  Naively speaking, at the transition point with
$z=2, \chi=0$, $v_m$ is irrelevant at $d=2$ for $m>2$.  However, this does
not necessarily imply irrelevance of the $m$-overlaps in the single chain
problem. Remember, that $v_m$ is a coupling introduced by hand in the many
chain Hamiltonian in Eq. \ref{eq:mpoly} to calculate the overlaps and {\it
does not} appear in the description of the single chain problem.  This opens
up a the possibility where under special conditions higher order overlaps
can become important as in multicritical cases or polymers with higher order
composite operators becoming relevant.  Proper choice of parameters can then
lead to multicritical analogs of spin glasses.  These are not the random
version of pure multicritical models but are of inherently different type.
In both cases, if the $m$-overlaps are important at the transition, they are
expected to be so in the spin glass phase also.  We hope that this will
motivate further detailed numerical work to settle this issue.  We conclude
that, like multifractals, a spin glass transition in finite dimensions
subsumes an infinite number of independent exponents.  This indicates either
a failure of the simple replica picture in finite dimensions or the
possibility of highly complex spin glass phases.

We thank A. Baumg\"artner for discussions and hospitality at KFA-IFF.  The
visit of the authors is supported by the Indo-German collaboration PHY-25/1.

\begin{figure}
\caption{ Diagrammatic representation of the parameters (a) and the
  solution (b). (See also Figs 1 and 2 of Ref. 5b.)  Solid square represents
  the vertex function which for zero external momenta gives $v_{mR}$.  The
  dotted line in the series for $v_m$ is a dummy line signifying loop
  closing. This line indicates that two different indices are coupled by the
  dummy momentum $p$. There are two factors in (b). (i) A combinatorial
  factor 8 from the insertions of $\lambda$ vertices and the subsequent
  noise contraction, and (ii) $(_2^m)$ for choosing the wave vectors.}
\end{figure}

\begin{references}
\bibitem{comm0} Henceforth, we use ``overlap'' to refer to 2-overlap
  and reserve the term ``$m$-overlap'' to overlaps for $m>2$.
\bibitem{comm2} Specific heat calculation gives $\nu \sim 6.7$
  \cite{der1}(a) while size calculations give $ \nu=4.3$
  \cite{der1}(b) for d=3. This needs to be sorted out.  In any case,
  the precise values of the single chain exponents are not so
  important for our discussion.
\end{references}
\end{document}